\documentclass[twocolumn,showpacs,preprintnumbers,amsmath,amssymb]{revtex4}
\usepackage{graphicx}
\usepackage{bm} %bold math
\usepackage[]{amsfonts, graphicx, amsmath}
\DeclareGraphicsExtensions{.eps,.ps,.eps.gz,.ps.gz,.eps.Z}

\newcommand{\ra}{\rangle}
\newcommand{\la}{\langle}
\newcommand{\tr}{{\rm Tr}}
\newcommand{\be}{\begin{equation}}
\newcommand{\ee}{\end{equation}}
\newcommand{\ber}{\begin{eqnarray}}
\newcommand{\eer}{\end{eqnarray}}

\newtheorem{theorem}{Theorem}
\newtheorem{lemma}{Lemma}

\newenvironment{proof}[1][Proof]{\begin{trivlist}
\item[\hskip \labelsep {\bfseries #1}]}{\end{trivlist}}
\newenvironment{definition}[1][Definition]{\begin{trivlist}
\item[\hskip \labelsep {\bfseries #1}]}{\end{trivlist}}

\newcommand{\qed}{\nobreak \ifvmode \relax \else
      \ifdim\lastskip<1.5em \hskip-\lastskip
      \hskip1.5em plus0em minus0.5em \fi \nobreak
      \vrule height0.75em width0.5em depth0.25em\fi}

\begin{document}

\title{Local purity distillation with bounded classical communication}
\author{\sc Hari Krovi}\email{krovi@usc.edu}
\author{Igor Devetak}%\email{devetak@usc.edu}
\affiliation{Communication Sciences Institute, University of Southern California, \\
Los Angeles, California 90089, USA}
%\subjclass{}
%\keywords{}
%\thanks{}
\date{\today}

\begin{abstract}
Local pure states are an important resource for quantum computing. The problem of distilling local pure states from mixed ones can be cast in an information theoretic paradigm. The bipartite version of this problem where local purity must be distilled from an arbitrary quantum state shared between two parties, Alice and Bob, is closely related to the problem of separating quantum and classical correlations in the state and in particular, to a measure of classical correlations called the one-way distillable common randomness. In Phys. Rev. A 71, 062303 (2005), the optimal rate of local purity distillation is derived when many copies of a bipartite quantum state are shared between Alice and Bob, and the parties are allowed unlimited use of a unidirectional dephasing channel. In the present paper, we extend this result to the setting in which the use of the channel is bounded. We demonstrate that in the case of a classical-quantum system, the expression for the local purity distilled is efficiently computable and provide examples with their tradeoff curves.
\end{abstract}


\maketitle

\section{Introduction}
One important goal of quantum information theory is to determine the optimal way of performing information theoretical tasks given information processing resources such as quantum or classical channels (both noisy and noiseless), shared entanglement and common randomness. The optimal protocol is usually defined as one that makes minimal use of the resources in the problem (in an asymptotic sense) to achieve the given information theoretic task. Often we assume a paradigm in which some resources are given for free. For example, one common paradigm is LOCC (local operations and classical communication), in which Alice and Bob can communicate classically at no cost. Local resources are also taken for granted in this paradigm, i.e., Alice and Bob have unlimited access to local quantum operations and local pure state ancillas. However, the problem of producing these local resources is of equal importance for various reasons. Local pure states are valuable as a computational resource since many computations require pure state ancillas to maintain unitarity. They are often used in protocols to perform information theoretic tasks. They are also useful in a thermodynamic sense where a pure state can be regarded as fuel to perform work \cite{L61}. In this paper, we consider the problem of distilling local pure states from bipartite quantum states.

The problem of distilling local pure states is first considered in \cite{OHHH} and \cite{HHHHOSS}, where concentrating purity from a quantum state was introduced. The quantum state can be localized to a single party or it can be distributed between two parties. Thus there are, broadly, two versions of this problem--the local version ({\it purity concentration}) and the distributed version ({\it purity distillation}). In the first one, Alice has $n$ copies of a quantum state $\rho^A$ and she needs to extract purity from it by performing only unitary operations. It has been shown in \cite{HHO} that the maximum possible asymptotic rate in the limit where $n$ tends to infinity is the difference between the number of qubits used to decribe the system and the von Neumann entropy of the system. In the second version, Alice and Bob share a large number of copies of a bipartite quantum system $\rho^{AB}$ and they need to distill local pure states using only local unitaries (instead of any local operation) and a dephasing channel which corresponds to classical communication \cite{OHHH}. Some bounds on this problem with one-way and two-way dephasing channels have been obtained in \cite{HHHHOSS}. In \cite{OHHHH}, the complementarity between the local information (which can be concentrated into local purity) and non-local information has been demonstrated i.e., the two parties can gain access to only one kind of information but not both. In \cite{D04}, both versions, the local and distributed, have been addressed in detail. The optimal one-way local purity distillation rate for the distributed case has been determined in terms of information theoretic quantities. This was related to an operational measure of classical correlations, the one-way distillable common randomness \cite{DW03}. This seems to indicate that the problem of distilling local purity is closely related to characterizing correlations in bipartite quantum states.

The problem of separating quantum and classical correlations is first considered in \cite{HV01}. In \cite{OHHH}, a measure of classical correlations is introduced: the so called {\it classical deficit} $\Delta^c_{\rightarrow}(\rho^{AB})$ and an operational approach to quantifying quantum correlations is given. In this problem, Alice and Bob share a state and have access to local heat baths. Using only LOCC, they have to concentrate the information in a state so that it can be used to do work. Indeed, as Landauer \cite{L61} noted, erasing information and resetting the system to a pure state requires work. Conversely, having pure states is a resource from which useful work can, in principle, be extracted. Now, if Alice and Bob can perform global operations, then the amount of work that can be done is more than when they use LOCC. The difference is represented by the deficit and corresponds to purely quantum correlations. Thus the problem of purity distillation is clearly related to the problem of characterizing quantum correlations. The above thermodynamical approach can been translated to an information theoretic approach \cite{HHHHOSS}. In this setting, the problem of distilling local pure states can be related to the problem of distilling common randomness from a bipartite quantum state \cite{DW03}. The ``regularized" version of the deficit $D_{\rightarrow}(\rho^{AB})$ is given an operational meaning in \cite{DW03} and it was shown to be equal to the {\it one-way distillable common randomness} or the 1-DCR. We extend this correspondence to the case when the rate of the one-way classical communication (represented here by the dephasing channel) is limited. We see that in the case of classical-quantum systems shared by Alice and Bob, the capacity obtained here is still similar to the 1-DCR.

Here we consider the problem of distilling local pure states from a bipartite quantum state using only local unitary operations and limited use of a one-way dephasing channel. In addition, the two parties, Alice and Bob, can borrow pure state ancillas from the environment as long as they are returned after the protocol. In this setting, we obtain the optimal tradeoff between the rate of purity distilled and the classical communication rate in terms of information theoretic quantities. This paper is organised as follows. In Sec. II, we describe the notation and some basic definitions. In Sec. III, we state and prove our main result. In Sec. IV, we show that the capacity becomes efficiently computable for the special case of classical-quantum systems i.e., when the joint state can be described as an ensemble of which Alice holds the classical part. We provide examples and their tradeoff curves. Finally, in Sec. V, we present our conclusions.

\section{Notation}
A quantum system $B$ is represented as an ensemble of quantum states $\mathcal{E}=(p(x),\rho^B_x)_{x\in\mathcal{X}}$ when the system $B$ is in the state $\rho_x$ with a probability $p(x)$. This can be represented equivalently by a {\it classical-quantum} system $XB$ as the state
\begin{equation}
\rho^{XB}=\sum_{x\in\mathcal{X}} p(x) |x\ra\la x|^X\otimes \rho_X^B ,
\end{equation}
where $\mathcal{H}_X$ has a preferred orthonormal basis $\{|x\ra\}_{x\in\mathcal{X}}$. One can see that the state $\rho^X=\sum_x p(x) |x\ra\la x|$ is equivalent to a random variable with distribution $p$. The state of $B$ is $\rho^B=\tr_X\rho^{XB}$ and we call $\rho^{XB}$ the {\it extension} of $\rho^B$. A pure extension i.e., when $\rho^{XB}$ is in a pure state, is called a {\it purification}. Given a bipartite quantum system $\rho^{AB}$, one can obtain the ensemble $\mathcal{E}$ by performing a POVM $\Lambda=(\Lambda_x)_x$ on the $A$ part of the system which makes $p(x)=\tr(\Lambda_x\rho^A)$ and $\rho^B_x=p(x)^{-1}\tr_A((\Lambda_x^A\otimes I^B)\rho^{AB})$. This can be thought of as a quantum map $\Lambda : \mathcal{H}_{A}\rightarrow \mathcal{H}_{X}$. Any classical map $f:\mathcal{X}\rightarrow\mathcal{Y}$ can be made a quantum map as
\begin{equation}\label{qmap}
f(\rho)=\sum_{x\in\mathcal{X}} \la x|\rho |x\ra |f(x)\ra\la f(x)|^Y,
\end{equation}
where $\{|y\ra\}_{y\in\mathcal{Y}}$ is the preferred basis for $\mathcal{H}_Y$.

The von Neumann entropy of a quantum state $\rho^A$, written as $H(A)_\rho$, is defined as $H(A)_\rho=-\tr(\rho^A\log \rho^A)$. For a classical-quantum system $XB$, the von Neumann entropy of the system $X$ is just the Shannon entropy of the random variable $X$ whose distribution is $p(x)$, i.e., $H(X)=-\tr(\rho^X \log \rho^X)=-\sum_x p(x) \log p(x)$. The conditional entropy is defined as
\begin{equation}
H(A|B)=H(AB)-H(B) .\nonumber
\end{equation}
The mutual information is
\[
I(A;B)=H(A)+H(B)-H(AB) .
\]
The distance measure used in this paper is the trace norm of an operator defined as
\[
||\omega ||_1=\tr \sqrt{\omega^\dag\omega} .
\]
Two states are said to be $\epsilon$-close if $||\rho-\sigma ||_1\leq\epsilon$. A sequence of symbols $y_1,y_2,\dots y_n$ is denoted as $y^n$ in this paper and a state $\rho_{y_1}\otimes\rho_{y_2}\otimes\dots\otimes\rho_{y_n}$ as $\rho_{y^n}$. Finally, a generic pure state for a system $A$ is represented as $|0\ra^A$.

\section{Local purity distillation from a quantum state}
Let us first review a result on the maximum rate of purity that can be distilled from many copies of a quantum state possessed by a single party. Let there be $n$ copies of a state $\rho^A$ defined on a system $A$ of dimension $d_A$. An $(n,P,\epsilon)$ purity concentration protocol is defined as a unitary operation $U:\mathcal{H}_{A^n}\rightarrow \mathcal{H}_{A_p}\otimes \mathcal{H}_{A_g}$ such that if $\Upsilon^{A_pA_g}=U\rho^{\otimes n}U^\dag$,
\begin{equation}
||\Upsilon^{A_p}-|0\ra\la 0|^{A_p}||_1\leq \epsilon .\nonumber
\end{equation}
The rate of this code is defined as $P=\frac{1}{n} \log d_{A_p}$, where $d_{A_p}=$ dim $\mathcal{H}_{A_p}$. A rate $P$ is said to be achievable if for all $\epsilon ,\delta > 0$ and sufficiently large $n$ there exists an $(n,P-\delta,\epsilon)$ protocol. The purity $\kappa (\rho)$ is defined as the supremum over all achievable rates. The following lemma is proved in \cite{HHHHOSS} and \cite{D04}.

\begin{lemma}[Purity concentration]\label{local_scenario}
The purity of the state $\rho^A$ of a quantum system $A$ of dimension $d_A$ is given by
\begin{equation}
\kappa(\rho^A)=\log d_A - H(A)_\rho .\nonumber
\end{equation}
\end{lemma}
This shows that the purity present in the state is complementary to the information content of the state as represented by the von Neumann entropy.

In this paper, we are concerned with the case when the initial state is distributed between Alice and Bob. In this scenario, Alice and Bob share many copies of a quantum state $\rho^{AB}$. They are allowed any local unitary operation and are in possession of a dephasing channel going from Alice to Bob. A dephasing channel is defined as a map $\mathcal{P}:\mathcal{H}_X\rightarrow\mathcal{H}_X$
\[
\mathcal{P}(\rho)=\sum_x |x\ra\la x|\rho|x\ra\la x|,
\]
where $\{|x\ra\}$ is an orthonormal basis for $\mathcal{H}_X$. It is equivalent to a classical communication channel (cf. Eq. (\ref{qmap}) with $f(x)=x$). This paradigm is an extension of LOCC and is referred to as 1-CLOCC$^\prime$. Closed LOCC or CLOCC is a paradigm in which Alice and Bob are not given unlimited access to pure state ancillas. In CLOCC$^\prime$, Alice and Bob are allowed ancillas as long as they return them at the end of the protocol. Such a protocol is called {\it catalytic}. Finally in 1-CLOCC$^\prime$, bidirectional classical communication is replaced by one-way classical communication from Alice to Bob. In this paper, the rate of this communication is bounded. We begin by assuming that we are given $n$ copies of an initial quantum state of Alice and Bob ($\rho^{AB}$) and another quantum system $C$ (which acts as a catalyst) of dimension $d_C$. An $(n,R,P,\epsilon)$ local purity distillation protocol comprises:
\begin{itemize}
\item a unitary operation on Alice's system $U_A$ : $\mathcal{H}_{A^n}\otimes\mathcal{H}_C\rightarrow\mathcal{H}_{A_p}\otimes\mathcal{H}_Y\otimes\mathcal{H}_{A_g}$.
\item a dephasing channel $\mathcal{P}$ : $\mathcal{H}_Y\rightarrow\mathcal{H}_Y$ from Alice to Bob.
\item a unitary operation on Bob's system $U_B$ : $\mathcal{H}_{B^n}\otimes\mathcal{H}_Y\rightarrow\mathcal{H}_{B_p}\otimes\mathcal{H}_{B_g}$,
\end{itemize}
such that, for $\frac{1}{n}\log d_Y\leq R$ we have
\begin{equation}\label{purity_trace}
\left\|\Upsilon^{A_pB_p}-|0\ra\la 0|^{A_p}\otimes |0\ra\la 0|^{B_p}\right\|_1\leq\epsilon ,
\end{equation}
where
\begin{equation}\label{purity_condition}
\Upsilon^{A_pA_gB_pB_g}=U_B\circ\mathcal{P}\circ U_A((\rho^{AB})^n\otimes |0\ra\la 0|^C).
\end{equation}
The rate of this code is defined as $P=\frac{1}{n}(\log d_{A_pB_p}-\log d_C)$, since the catalyst rate of $\frac{1}{n}\log d_C$ must be returned from the distilled purity. A rate pair $(P,R)$ is said to be achievable if for all $\epsilon$, $\delta > 0$ and sufficiently large $n$, there exists an $(n,R+\delta,P-\delta,\epsilon)$ protocol. The {\it 1-way distillable local purity} $\kappa_{\rightarrow}(\rho^{AB},R)$ is defined as the supremum of $P$ over all achievable rates pairs $(P,R)$. We are now ready to state the main theorem.

\begin{theorem}\label{main_theorem}
The 1-way distillable local purity of the state $\rho^{AB}$ is given by $\kappa_\rightarrow=\kappa^\ast_\rightarrow$, where
\begin{equation}\label{k_star}
\kappa^\ast_{\rightarrow}(\rho^{AB},R)=\kappa(\rho^A)+\kappa(\rho^B) + P_\rightarrow (\rho^{AB},R) ,
\end{equation}
\begin{equation}
P_\rightarrow (\rho^{AB},R)=\lim_{n\rightarrow\infty} \frac{1}{n} P^{(1)}_\rightarrow ((\rho^{AB})^{\otimes n},nR) ,\nonumber
\end{equation}
\begin{equation}\label{P}
P^{(1)}_\rightarrow (\rho^{AB},R)=\max_\Lambda \{I(Y;B)_\sigma : I(Y;BE)_\sigma\leq R\} 
\end{equation}
and
\be\label{sigma}
\sigma^{YBE}=(\Lambda\otimes I^{BE})(\Psi^{ABE}) .
\ee
$\Psi^{ABE}$ is a purification of $\rho^{AB}$, $\Lambda$ is a POVM on Alice's system and the maximization is over all POVMs $\Lambda : \mathcal{H}_A\rightarrow \mathcal{H}_Y$.
\end{theorem}

The purity $\kappa_{\rightarrow}(\rho^{AB},R)$ contains contributions of two kinds. The term $\kappa(\rho^A)+\kappa(\rho^B)$ reflects the amount of purity that Alice and Bob can distill individually using the unitary from Lemma \ref{local_scenario}. The term $P_\rightarrow (\rho^{AB},R)$ reflects the purity that can be distilled by using the fact that Alice and Bob share a quantum system i.e., they are correlated. We note that the expression for $\kappa_{\rightarrow}(\rho^{AB},R)$ above would coincide with the expression for purity in \cite{D04}  if we replace $P_\rightarrow (\rho^{AB},R)$ with $D_\rightarrow (\rho^{AB})$, the 1-DCR. We show later, that when $R\rightarrow\infty$, the two quantities are the same.

The proof of the theorem consists of two parts: the direct coding theorem and the converse. The direct coding theorem establishes that every $P$ such that $P\leq \kappa^\ast_{\rightarrow} (\rho^{AB},R)$ corresponds to a protocol with an achievable rate pair $(P,R)$. The converse establishes that if the rate pair $(P,R)$ is achievable then $P\leq \kappa^\ast_{\rightarrow} (\rho^{AB},R)$.
\begin{proof}[Proof of theorem \ref{main_theorem}: Converse.]
Consider a general $(n,R,P,\epsilon)$ purity distillation protocol and let $\delta=\frac{2}{en} + \epsilon\log d_Ad_B$. We have that
\begin{equation}
\log d_{A_p} d_{B_p} - \log d_C=\log d_{A^nB^n} -\log d_{A_gB_g} .
\end{equation}
 We can bound the dimension of the ``impure" part of the state obtained, i.e. $A_gB_g$, through a set of inequalities in the following way (the explanations are given below). 
 \begin{eqnarray}
\log d_{A_g}d_{B_g} &\geq& H(B_g) +H(A_g) \label{1} \\
&\geq& H(B_pB_g)-H(B_p)+H(A_g) \label{2} \\
&=& H(B^nY') - H(B_p)+H(A_g) \label{3} \\
&\geq& H(B^nY)-H(B_p) + H(A_g) \label{4} \\
&=& H(B^n|Y)+H(Y)+H(A_g)-H(A_p) \nonumber \\
&-&H(B_p)+H(A_p)\label{5} \\
&\geq& H(B^n|Y)+ H(YA_pA_g)- H(A_p) \nonumber \\
&-& H(B_p)\label{6} \\
&\geq& H(B^n|Y)+ H(A^nC)- H(A_p)\nonumber \\
&-& H(B_p)\label{7} \\
&\geq& H(B^n|Y)+H(A^n) - n\delta \label{8} \\
&\geq& nH(A)+nH(B)-I(Y ; B^n)-n\delta .\label{9}
\end{eqnarray}
Eq. (\ref{1}) follows from the fact that the maximum of $H(\rho^S)$ of any state $\rho^S$ is $\log d_S$, where $d_S$ is the dimension of the Hilbert space of $\rho^S$. The second line Eq. (\ref{2}) follows from the fact that conditioning does not increase entropy (see Eq. (\ref{A_cond}) in the appendix) and the definition of conditional entropy given in Sec. II. In Eq. (\ref{3}), $Y'$ denotes the system $Y$ after dephasing. Here we use the fact that unitary operations do not change the total entropy i.e., $H(B_pB_g)=H(B^nY')$. In Eq. (\ref{4}), we use the fact that dephasing cannot decrease entropy \cite{NC01} and thus replace $Y'$ by $Y$. In Eq. (\ref{5}), we use the definition of conditional entropy again and in Eq. (\ref{6}), we use the subadditivity of entropy (see Eq. (\ref{A_sa}) in the appendix) on $H(Y)+H(A_g)+H(A_p)$. Eq. (\ref{7}) again uses the fact that unitary operations do not change the total entropy i.e., $H(YA_pA_g)=H(A^nC)$. Eq. (\ref{8}) follows from Fannes' inequality and the condition in Eq. (\ref{purity_trace}). Eq. (\ref{9}) uses the definition of quantum mutual information given in Sec. II. Thus we have
\begin{eqnarray}\label{P_eqn}
P &=& \frac{1}{n} (\log d_{A_pB_p}-\log d_C) \nonumber \\
&\leq& \log d_Ad_B - H(A)-H(B)+ \frac{1}{n} I (Y ; B^n) +\delta \nonumber \\
&=&\kappa(\rho^A)+\kappa(\rho^B)+ \frac{1}{n} I (Y ; B^n) +\delta .
\end{eqnarray}
Let $E$ be the system that purifies the initial state $\rho^{AB}$. The classical communication rate is bounded by
\be
R=\frac{1}{n}\log d_{Y} \geq \frac{1}{n}H(Y)\geq \frac{1}{n}I(Y;B^nE^n) .
\ee
It now easily follows that if the rate pair $(P,R)$ is achievable, then $P\leq \kappa_{\rightarrow}(\rho^{AB},R)$.
%The sequence of inequalities above show that for any possible protocol, we have
%\ber
%P &\leq& \log d_A + \log d_B -H(A)-H(B) + \frac{1}{n} I (Y^n ; B^n) +\delta \nonumber \\
%&\leq& \kappa(\rho^A) + \kappa(\rho^B) + \frac{1}{n} P_{\rightarrow}^{(1)}((\rho^{AB})^{\otimes n},R) \label{P2} \\
%&\leq& \kappa(\rho^A) + \kappa(\rho^B) + \lim_{n\rightarrow\infty} \frac{1}{n} P_{\rightarrow}^{(1)}((\rho^{AB})^{\otimes n},R) \label{P3} \\
%&=& \kappa_{\rightarrow}(\rho^{AB},R) \nonumber .
%\eer
%Eq. (\ref{P2}) follows from the maximization taken over all measurements. Eq. (\ref{P3}) is a consequence of the fact that $P_n(\rho,R)=\frac{1}{n} P_{\rightarrow}^{(1)}((\rho^{AB})^{\otimes n},R)$ is monotonically increasing. This can be shown by using the relation between $P_n(\rho,R)$ and $D_n(\rho,R)$ derived in Sec. IV. It has been shown in \cite{DW03} that $D_n(\rho,R)$ is monotonically increasing. Therefore, it follows that $P_n(\rho,R)$ is also monotonically increasing. This proves the converse.
\end{proof}

Before we prove the direct coding theorem, we need the following lemmas. The first of these comes from the measurement compression theorem proved in \cite{W01}.
\begin{lemma}\label{winter_meas}
Let $\rho^{AB}$ be a quantum state with a purification $\Psi^{ABE}$, and  $\Lambda=(\Lambda_k)_k$ a POVM, where $\Lambda : \mathcal{H}_{A}\rightarrow \mathcal{H}_{Y}$ and $\mathcal{Y}$ is the set of outcomes of $\Lambda$. For any $\epsilon$, $\delta$ $>0$ and all sufficiently large $n$, there  exist a set of POVMs $(\tilde{\Lambda}_m ^{(l)})_m : \mathcal{H}_{A^n}\rightarrow \mathcal{H}_{Y^n}$, whose outcomes $m$ lie in a set $\mathcal{M}$, where $|\mathcal{M}| = 2^{(nI(Y;BE)+n\delta)}$ and are indexed by $l\in \mathcal{L}$ where $|\mathcal{L}|= 2^ {(nH(Y|BE)+n\delta)}$, and a deterministic function $f:\mathcal{M}\times \mathcal{L}\rightarrow \mathcal{Y}^n$ such that for 
\be
\tilde{\Lambda} : \sigma^{A^n} \rightarrow \frac{1}{|\mathcal{L}|} \sum_l \tr(\tilde{\Lambda}_m^{(l)}\sigma^{A^n}) |f(m,l)\ra\la f(m,l)|^{Y^n},
\ee
we have
\[
||({\rm id}^{B^nE^n}\otimes \tilde{\Lambda})((\Psi^{ABE})^{\otimes n})-(({\rm id}^{BE}\otimes \Lambda)(\Psi^{ABE}))^{\otimes n}||_1 \leq \epsilon .
\]
\end{lemma}
The lemma states that we can perform a block measurement $\tilde{\Lambda}$ instead of a tensor product measurement $\Lambda^{\otimes n}$ on $n$ copies of a state such that the outcome states are $\epsilon$-close. The main idea is that one can compress the information obtained by $\Lambda^{\otimes n}$ which lies in a larger set to the outcomes of $\tilde{\Lambda}^{(l)}$ which lie in a smaller set. When we average over all the possible values of $l$, the states obtained by the two measurements are close.

We now give the definition and a result related to typical sets used in classical information theory \cite{CovTho}.
\begin{definition}\label{typical_def}
Given a random variable $Y$ with a probability distribution $p$, a $\delta$-typical set is defined as
\be
\mathcal{T}_{Y,\delta}^n=\{y^n:\forall y , |n_y-np_y|\leq \delta n\} ,
\ee
where $n_y$ is the number of occurrences of $y$ in $y^n$ and $p_y$ is the probability of the symbol $y$.
\end{definition}

\begin{lemma}\label{typical}
For any $\epsilon$, $\delta$ $>0$ and all sufficiently large $n$, ${\rm Pr}(y^n\in\mathcal{T}_{Y,\delta}^n) \geq 1-\epsilon$.
\end{lemma}

\begin{proof}[Proof of theorem 1: Direct coding.]
The direct coding theorem involves a protocol which distills purity in two ways. Firstly, there is the local purity of Alice and Bob which gives rise to the $\kappa (\rho^A)$ and $\kappa (\rho^B)$ terms from Eq. (\ref{k_star}). Secondly, classical correlations between them allow for extra purity to be distilled which gives the $P_{\rightarrow}(\rho^{AB},R)$ term. The main idea is to use redundant information of a party to erase a part of it and thus create purity. We use the measurement compression theorem from lemma \ref{winter_meas}, to substitute an arbitrary measurement on $n$ copies of the initial state by block measurements on these $n$ copies with a smaller number of outcomes, but which extract the same information. This is useful in deducing the classical communication required for the protocol.

 Alice is initially in possession of a system in the state $(\rho^A)^{\otimes n}$ i.e., $n$ copies of a state $\rho^A$. Let $\Lambda : \mathcal{H}_{A}\rightarrow \mathcal{H}_Y$ be a POVM on Alice's system. Assume that Alice and Bob share common randomness of size $|\mathcal{L}|$. We later show (at the end of this section) that the common randomness is not needed. Now, using lemma \ref{winter_meas} the measurement $\Lambda^{\otimes n}$ can be compressed to $\tilde{\Lambda} : \mathcal{H}_{A^n}\rightarrow\mathcal{H}_{Y^n}$. The outcomes lie in a set of size $|\mathcal{M}|$. The basic steps of the protocol are as follows: Alice performs the measurement $\tilde{\Lambda}$ {\it coherently} i.e., in a unitary fashion by copying the outcomes onto a borrowed quantum system $M$ of size $|\mathcal{M}|$, originally in a pure state. She then performs purity concentration (lemma \ref{local_scenario}) on her $A^n$ system conditioned on the content of $M$ and sends the system $M$ to Bob through the dephasing channel. Bob performs purity concentration on his state $B^n$ after receiving $M$, again conditioned on the content of $M$. 
 
 We now analyze each of these steps to determine the amount of purity that we obtain. We do this by first analyzing two easier hypothetical scenarios and then apply the results to the actual protocol. 
 \begin{enumerate}
 \item Let us first assume that Alice borrows the system $Y^n$ (not $M$) of size $|\mathcal{Y}|^n$ in a pure state and performs $\Lambda^{\otimes n}$ instead of $\tilde{\Lambda}$. Moreover, she performs $\Lambda^{\otimes n}$ coherently using the ancilla. In other words, Alice could perform some unitary on the system $A^nY^n$ and then completely dephase the system $Y^n$ in a fixed basis $\{|y^n\ra\}$. But since we send the state through the dephasing channel (which dephases in the basis $|y^n\ra$), if we show that the unitaries performed and the channel commute, we can assume that Alice performs the actual measurement instead of a coherent version of it. We show below that this is indeed the case i.e., we derive a form for the unitary operation that Alice performs which clearly commutes with the dephasing channel. Therefore, we assume now that Alice has the post measurement state. If Alice performs $(\Lambda)^{\otimes n}$ on her state $\rho^{A^n}$, she obtains an ensemble $\{p(y^n),\rho_{y^n}\}$ and her combined state will be
 \be
 \omega^{A^nY}=\sum_{y^n} p(y^n) \rho_{y^n} ^{A^n}\otimes |y^n\ra\la y^n|^{Y} .
 \ee
 Now we choose $\epsilon$, $\delta$ and sufficiently large $n$ so that there exist $(n,H(A)_{\rho_y}-\delta,\epsilon)$ protocols for each $y$. Alice can transform a state from $\rho_{y^n}$ to $\bigotimes_y \rho_y^{\otimes n_y}$ by performing a local unitary operation for each $y^n$ i.e., a conditional unitary, since the two states are related by a permutation of Alice's input Hilbert spaces. From the definition of a typical set above, when $y^n\in \mathcal{T}^n_{Y,\delta}$, then $np_y-n\delta \leq n_y\leq np_y+n\delta$ and if $n$ is sufficiently large, $p(y^n\notin \mathcal{T}^n_{Y,\delta})\leq \epsilon$ by lemma \ref{typical}. If $y^n$ lies in the typical set, we can construct a conditional unitary that permutes Alice's input Hilbert spaces in the following way:
 \be
 y^n \rightarrow \underbrace{y_1\dots y_1}_{n(p_{y_1}- \delta)}\underbrace{y_2\dots y_2}_{n(p_{y_2}- \delta)}\dots\underbrace{y_k\dots y_k}_{n(p_{y_k}- \delta)} y_g ,
 \ee
 where $y_g$ denotes the remaining symbols in $y^n$ in some order and $k=|\mathcal{Y}|$. Thus, by a unitary transformation we obtain the state $\bigotimes_y \rho_y^{\otimes n(p_y-\delta)}\otimes \rho_{y_g}$, where ${\rm dim}(\rho_{y_g})\leq 2k\delta \log d_A$. Now, Alice can perform purity concentration using lemma \ref{local_scenario} on her new state. In order to purify the state $\rho_{y_1}^{\otimes n(p_{y_1}-\delta)}$ Alice performs a fixed unitary $U_1$ given by lemma \ref{local_scenario} and obtains a purity of $n(p_y-\delta)(\log d_A - H(\rho_{y_1})-\delta)$. Similarly, to purify $\bigotimes_y \rho_y^{\otimes n(p_y-\delta)}\otimes \rho_{y_g}$, Alice performs a tensor product of fixed unitary operations $\bigotimes_iU_i\otimes I$, where $I$ is the identity operator on the $\rho_{y_g}$ system, and distills purity of rate 
 \ber
 P_A&=& \sum_y n(p_y-\delta)(\log d_A)\nonumber \\
 &-& \sum_y n(p_y-\delta) (H(\rho_y)-\delta) \nonumber \\
% &=& (n-\sum_y n\delta)\log d_A - nH(A|Y) + n\delta \sum_yH(\rho_y) \nonumber \\
 &=& n(\log d_A - H(BE|Y))_\sigma - n\delta_A,
 \eer
 where $\delta_A=(k\log d_A-\sum_y H(\rho_y)+1)\delta$. Recall from Eq. (\ref{sigma}) that $\sigma^{YBE}=(\Lambda\otimes I^{BE})(\Psi^{ABE})$. This operation leaves the state $\bigotimes_y \rho_y^{\otimes n(p_y-\delta)}\otimes \rho_{y_g}$ in a state $k \epsilon$-close to $|0\ra\la 0|^{nP_A}\otimes \rho_{y^n}^{\prime A_g}$, where $\rho_{y^n}^{\prime A_g}$ is some resultant state on the system $A_g$ and will be discarded eventually. But Alice's state is given by
 \ber
 \omega^{A^nY}&=&\sum_{y^n} p(y^n)\rho^{A^n}_{y^n}\otimes |y^n\ra\la y^n|^Y\nonumber \\
 &=&\sum_{y^n\in\mathcal{T}^n_{Y,\delta}}p(y^n)\rho^{A^n}_{y^n}\otimes |y^n\ra\la y^n|^Y \nonumber \\
 &+& \sum_{y^n\notin\mathcal{T}^n_{Y,\delta}}p(y^n)\rho^{A^n}_{y^n}\otimes |y^n\ra\la y^n|^Y.
 \eer 
 When we apply the fixed unitary $\bigotimes_iU_i\otimes I$ to this state, we obtain a state $k \epsilon$-close to
 \ber
 |0\ra\la 0|^{nP_A} &\otimes& \sum_{y^n\in\mathcal{T}^n_{Y,\delta}}p(y^n)\rho_{y^n}^{\prime A_g} \nonumber \\
 &+& \sum_{y^n\notin\mathcal{T}^n_{Y,\delta}}p(y^n)(\bigotimes_iU_i\otimes I)\rho_{y^n}^{A^n} .
 \eer
 Since $p(y^n\notin \mathcal{T}^n_{Y,\delta})\leq\epsilon$, the trace of the second term in the above sum is small and by using Eq. (\ref{A_smtr}) in the appendix, we can see that the above state is $(k+2)\epsilon$-close to
 \be
 |0\ra\la 0|^{nP_A}\otimes \sum_{y^n\in\mathcal{T}^n_{Y,\delta}}p(y^n)\rho_{y^n}^{\prime A_g}.
 \ee
 It is useful to think of Alice's operations as a noisy map $\mathcal{E}$ which takes Alice's state $\omega$ to a state $(k+2)\epsilon$-close to $|0\ra^{nP_A}$. We can write $\mathcal{E}$ as the conditional unitary operation
 \be\label{U_form}
 \sum_{y^n} U_{y^n}^{A^n}\otimes |y^n\ra\la y^n|^{Y^n}
 \ee
 followed by discarding the $A_gY^n$ system i.e., if Alice's state were $\omega$, then
 \be
 \mathcal{E} : \omega \stackrel{(k+2)\epsilon}{\longrightarrow} |0\ra^{nP_A} ,
 \ee
 using obvious notation. We can also see from the form of this unitary in Eq. (\ref{U_form}) that if the dephasing channel dephases in the $|y^n\ra$ basis, it commutes with this unitary. Hence, we are justified in assuming that Alice can perform $\Lambda^{\otimes n}$ instead of the coherent version (we will show presently that this is also the case with $\tilde{\Lambda}$).

\item In this scenario, we assume that Alice performs $\tilde{\Lambda}$ after borrowing the system of size $|\mathcal{Y}|^n$ in a pure state, her state becomes
 \be
 \omega'=\sum_{y^n} \tilde{p}(y^n) \tilde{\rho}_{y^n} ^{A^n}\otimes |y^n\ra\la y^n|^{Y^n} ,
 \ee
By lemma \ref{winter_meas}, $\omega$ and $\omega'$ are $\epsilon$-close. Therefore, performing the same local unitaries to permute Alice's input Hilbert spaces and purity concentration would amount to a purity rate of
 \be
 \log d_A- H(BE|Y)_\sigma - \delta_A ,
 \ee
 and Alice's state would have been $(k+3)\epsilon$ close to $|0\ra^{nP_A}$ by the monotonicity of the trace distance under noisy maps (Eq. (\ref{A_mon}) in the appendix). Therefore, we would have
  \be
 \mathcal{E} : \omega' \stackrel{(k+3)\epsilon}{\longrightarrow} |0\ra^{nP_A} .
 \ee
 
\item But in reality, Alice and Bob share a common random index $l \in |\mathcal{L}|$ and Alice borrows system $M$ in a pure state. She performs the measurement given by $(\tilde{\Lambda}^{(l)})_m$ to obtain an ensemble $\{1/|\mathcal{L}|,\sigma_l\}$, where $\sigma_l$ is of the form
 \be
 \sigma_l=\sum_m p_m \rho_{l,m}^{A^n} \otimes |m\ra\la m|^M .
 \ee
 Instead of $\mathcal{E}$, Alice now can perform the map $\mathcal{E}_l$ which corresponds to the unitary operation
  \be
 \sum_{m} U_{f(m,l)}^{A^n}\otimes |m\ra\la m|^{M}
 \ee
 followed by discarding the $A_gM$ system. (From the form of this unitary, we can see that it would commute with a dephasing channel which dephases in the $|m\ra$ basis.) We now have that
 \be\label{e_l}
 \frac{1}{|\mathcal{L}|}\sum_l \mathcal{E}_l(\sigma_l) = \mathcal{E}(\omega') .
 \ee
 Therefore, the net effect is the same as if $\tilde{\Lambda}$ was performed on her state in scenario 2.
 \end{enumerate}
Alice now sends her $M$ system through the dephasing channel to Bob. Bob's state is an ensemble which can be written (similar to Alice's), conditioned on $M$ in the following way:
 \be\label{Bob_ML}
 \hat{\sigma}^{B^nM}_l=\sum_m p_m \hat{\rho}_{l,m}^{B^n} \otimes |m\ra\la m|^M .
 \ee
 Bob performs the same sort of permutation and purification operations as Alice to obtain a purity rate $P_B=n\log d_B-nH(B|Y)_\sigma-\delta_B$, where $\delta_B$ is proportional to $\delta$. Bob's state after these operations is $(k+3)\epsilon$ close to $|0\ra^{nP_B}$. The protocol uses a catalyst of size $ nI(Y;BE)_\sigma$ qubits and obtains a state $(k+3+k+3)\epsilon=(2k+6)\epsilon$ close to $|0\ra\la 0|^{nP}$. The purity rate $P$ is given by
\ber
P&=&\log d_A-H(BE|Y)_\sigma-I(Y;BE)_\sigma\nonumber \\
&+&\log d_B - H(B|Y)_\sigma -\delta_A-\delta_B \nonumber \\
&=& \log d_Ad_B - H(A)_\sigma -H(B)_\sigma+I(Y;B)_\sigma-\delta_A-\delta_B . \nonumber
\eer
The classical communication rate is $\frac{1}{n}\log |\mathcal{M}|= I(Y;BE)_\sigma$. Thus, the rate pair $(P,R)=(\kappa (\rho^A)+\kappa (\rho^B) +I(Y;B)_\sigma, I(Y;BE)_\sigma)$ is achievable. This proves the direct coding theorem using common randomness. 
We now show that the common randomness used in the protocol is not necessary through a process of derandomization. We can denote the performance of the whole protocol by the following map on the combined system of Alice and Bob. Let $\nu_l^{A^nB^nM}$ denote the state of the $A^nB^nM$ system after Alice borrows $M$ in a pure state and performs the measurements $(\tilde{\Lambda}^{(l)}_m)_m$. All the operations performed by Alice and Bob can be represented by a noisy map in the following way:
\be\label{ehat_l}
\hat{\mathcal{E}}_l : \nu_l^{A^nB^nM} \stackrel{(2k+6)\epsilon}{\longrightarrow} |0\ra\la 0|^{nP} .
\ee

 In Eq. (\ref{ehat_l}), let $\mu_l=\hat{\mathcal{E}}_l(\nu_l)$. We have that
\be
||\frac{1}{|\mathcal{L}|}\sum_l \mu_l -|0\ra\la 0| ||\leq \epsilon' ,
\ee
where $\epsilon'=(2k+6)\epsilon$. Using the relationship between the trace distance and fidelity in Eq. (\ref{A_ft}) in the appendix, we have that
\be
\frac{1}{|\mathcal{L}|}\sum_l \la 0|\mu_l|0\ra \leq 1- (\epsilon')^2/4.
\ee
This means that in the above average over the index $l$, there must exist some $l_0$ such that
\be
\la 0|\mu_{l_0}|0\ra \leq 1- (\epsilon')^2/4 .
\ee
Therefore, Alice and Bob can agree on this value $l_0$ for the protocol, use it to perform the measurement $\tilde{\Lambda}^{(l_0)}$ and obtain the same purity as with common randomness.
\end{proof}

\section{Examples}
In this section, we analyze the behavior of the purity rate for the special case of classical-quantum systems. We show that when Alice and Bob share a classical-quantum system, the purity rate becomes ``single-letterizable". More precisely, we show that the quantity 
\be
P_n(R)=\max_{Y|X^n}\{\frac{1}{n} I(Y;B^n) | \frac{1}{n}I(Y;B^nE^n)\leq R\}
\ee
satisfies $P_n(R)=P_1(R)$. When Alice and Bob share a classical-quantum system (denoted here as $\rho^{XB}$ instead of $\rho^{AB}$), a measurement on Alice's system becomes a classical channel $Y|X$, a conditional probability distribution.

Observe that $P_1(R)$ is exactly the same as $P_{\rightarrow}^{(1)}(\rho^{AB},R)$ defined in Eq. (\ref{P}) with a measurement replaced by the classical channel $Y|X$. We relate the above quantity to the quantity $D_n(R)$, defined in \cite{DW03} as
\be
D_n(R)=\max_{Y|X^n} \{\frac{1}{n} I(Y;B^n) | \frac{1}{n} (I(Y;X^n)-I(Y;B^n)\leq R\} .
\ee
Since the system $BE$ purifies $X$ and the system $Y$ is obtained from $X$ by passing it through the classical channel $Y|X$, we have that $I(Y;BE)=I(Y;X)$. Therefore, we arrive at the following relation between the quantities $P_n$ and $D_n$:
\be\label{PD_eq}
P_n(D_n(R)+R)=D_n(R) .
\ee
Therefore, we have
\be
P_n(R')=D_n(R'-D_n(R)),
\ee
where $R'=D_n(R)+R$. It has been shown in \cite{DW03} that $D_n(R)$ is single-letterizable and therefore, $R'=D_1 (R)+R$. Now
\be
P_n(R')=D_1 (R'-D_1 (R))=P_1 (R') .
\ee
Thus, it follows that $P_n(R)$ is also single-letterizable. 

\begin{figure}[h]
\begin{center}
\includegraphics[scale=0.3]{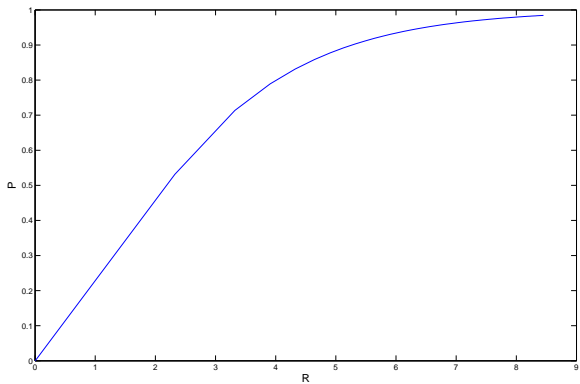}
\end{center}
\caption{$P(R)$ for the uniform ensemble.} \label{Uniform}
\end{figure}

We now consider two examples in this section: the uniform qubit ensemble and the parametrized BB84 ensemble \cite{DevBer,HJW}. The uniform qubit ensemble is a uniform distribution of pure states on the Bloch sphere. In this case, we obtain a curve shown in Fig. \ref{Uniform}. We can parametrize the purity rate and the communication rate in the following way using \cite{DevBer}, \cite{DW03} and Eq. (\ref{PD_eq}):
\ber
R&=&\frac{\lambda}{e^\lambda -1} - 1 + \log\left(\frac{\lambda e^\lambda}{e^\lambda -1}\right) \nonumber \\
P(R)&=& 1- h_2 \left(\frac{1}{\lambda} - \frac{1}{e^\lambda -1}\right) ,
\eer
for $\lambda\in (0,\infty)$ and where $h_2(p)=-p\log p - (1-p)\log (1-p)$ is the binary Shannon entropy. As Alice sends an increasing amount of information to Bob we have that $R\rightarrow\infty$. From the graph we can see that $P\rightarrow 1$ as $R\rightarrow\infty$. This means that eventually, Bob distills all the purity he can since he possesses a single qubit.

\begin{figure}[h]
\begin{center}
\includegraphics[scale=0.3]{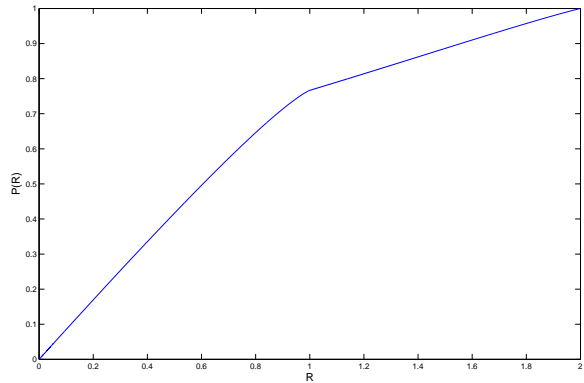}
\end{center}
\caption{$P(R)$ for the parametrized BB84 ensemble with $\theta=\frac{\pi}{8}$.} \label{BB84}
\end{figure}

The second example we consider is the parametrized BB84 ensemble $\mathcal{E}_{BB}(\theta)$ \cite{HJW}, which is defined as
\ber
|\phi_1\ra &=& |0\ra \nonumber \\
|\phi_2\ra &=& \cos\theta |0\ra + \sin\theta |1\ra \nonumber \\
|\phi_3\ra &=& |1\ra \nonumber \\
|\phi_4\ra &=& -\sin\theta |0\ra + \cos\theta |1\ra , \nonumber
\eer
each with a probability of $\frac{1}{4}$. Clearly, two classical bits are needed to specify the state. The local purity tradeoff curve is plotted in Fig. (\ref{BB84}). Note that $R=1$ is a special point on the curve at which the behavior changes. Alice's strategy to optimize purity for $R\leq 1$ is to consider the restricted ensemble consisting of $\frac{1}{2}(|\phi_1\ra\la \phi_1 |)+(|\phi_2\ra\la\phi_2 |)$ and $\frac{1}{2}(|\phi_3\ra\la\phi_3 |+|\phi_4\ra\la\phi_4|)$, where she ignores the bit that distinguishes between the states $|\phi_1\ra$ and $|\phi_2\ra$ and between $|\phi_3\ra$ and $|\phi_4\ra$. For $R>1$, the optimal strategy requires her to consider the full ensemble, and when $R=2$, Bob gets a full qubit of purity.

\section{Discussion}
We have discussed the problem of distilling local pure states from a bipartite quantum state shared between Alice and Bob when they have unlimited access only to local unitary operations. They can borrow local pure states in a catalytic sense, i.e., they must return them at the end of the protocol. Finally, they are allowed bounded classical communication at a rate that is upper bounded by $R$. We have derived an optimal protocol enabling them to distill local purity. We make use of common randomness in this protocol by assuming that Alice and Bob share $nH(Y|BE)$ random bits. However, we have derandomized the protocol to show that we do not really need this common randomness. The expression for the optimal rate of local purity does not seem to be single-letterizable in the general case, but it becomes single-letterizable when Alice and Bob have a classical-quantum system. We can now see that in the case of unlimited classical communication, $R\rightarrow\infty$, this reduces to the rate of local purity derived in \cite{D04}. Indeed the rate of classical communication in the protocol, i.e., $I(Y;BE)$ can be as large as we wish when $R\rightarrow\infty$. The maximum possible value of $I(Y;BE)$ is when $Y$ and $BE$ are perfectly correlated. In this case we get that $I(Y;BE)=H(BE)=H(A)$, which was the amount of classical communication needed to obtain the maximum local purity in \cite{D04}. This means that $P_\rightarrow (\rho^{AB},\infty)$ reduces to $\kappa_\rightarrow (\rho^{AB})$ obtained in \cite{D04}, where
\be
\kappa_{\rightarrow}(\rho^{AB})=\kappa(\rho^A)+\kappa(\rho^B)+\lim_{n\rightarrow\infty}\frac{1}{n} (\max_\Lambda I(X;B)_{(\Lambda\otimes I)\rho}).
\ee
The maximization above is over all rank-1 POVMs $\Lambda : \mathcal{H}_A\rightarrow \mathcal{H}_X$. In \cite{D04}, it was shown that $\kappa_\rightarrow (\rho^{AB})$ was exactly the one-way distillable common randomness. Our results in this paper show that this similarity can be extended even in the case of bounded classical communication for classical-quantum systems as given by Eq. (\ref{PD_eq}).

The measurement compression theorem \cite{W01} used in the protocol is a generalization of the classical reverse Shannon theorem \cite{BSST} and can be viewed as a quantum-classical channel simulation problem. Channel simulation is an essential ingredient in determining optimal tradeoffs for problems such as quantum data compression \cite{HJW}, entanglement distillation \cite{DHW05} and remote state preparation \cite{BHLSW}. Here we see another example of this rule of thumb.

In order to count local resources, the theory of resource inequalities \cite{DHW05} can be extended to include local resources. In \cite{D04}, a notation for local purity has been introduced and a unit of purity was called a {\it pbit}. It was represented as $[q]$ in a resource inequality. Using standard notation, a classical channel, a unit of common randomness and a bipartite quantum state are represented as $[c\rightarrow c]$, $[cc]$ and $\{qq\}$ respectively. If pure state ancillas are free, then a classical channel and a quantum dephasing channel are identical, but in the {\it closed} version of the protocol, we identify a dephasing channel with $[c\rightarrow c]$. The resource inequality for the protocol here, becomes
\be\label{resource}
\{qq\} + R [c\rightarrow c] \geq P_\rightarrow (\rho^{AB},R) [q] .
\ee

As further research, one could consider local purity distillation in the presence of  a two-way classical communication. But this problem, at least in the present setting, seems difficult. This may be because this problem is related to the common randomness distillation problem \cite{DW03}. The problem of distilling common randomness (as with entanglement distillation \cite{DW05}) becomes quite hard in the presence of two-way classical communication.

\section{Acknowledgments}
This work is supported in part by NSF grants CCF-0524811 and CCF-0545845 (CAREER).

\section{Appendix}
In this appendix, we collect some miscellaneous identities and inequalities used in the protocol. Most of these can be found in \cite{NC01}
\begin{enumerate}
\item Given states $\rho$ and $\sigma$, we have the following equality
\be\label{A_ui}
||\rho-\sigma ||_1= ||U\rho U^\dag- U\sigma U^\dag ||_1,
\ee
for any unitary $U$.
\item For states $\rho$, $\sigma$ and any completely positive trace preserving (CPTP) map $\mathcal{E}$, we have the inequality
\be\label{A_mon}
||\rho-\sigma ||_1\geq ||\mathcal{E}(\rho)-\mathcal{E}(\sigma) ||_1.
\ee
\item For any states $\rho$ and $\sigma$ and any $\epsilon>0$, we have
\be\label{A_smtr}
||((1-\epsilon)\rho + \epsilon\sigma) - \rho||_1=\epsilon ||\sigma -\rho ||_1\leq 2\epsilon.
\ee
\item The trace distance and fidelity are related by the following equation
\be\label{A_ft}
||\rho-|\phi\ra\la\phi | ||_1 \leq 2\sqrt{1-\la\phi|\rho|\phi\ra} .
\ee
\item Given $\rho$ and $\omega$ which are two states in a certain $d$-dimensional Hilbert space, we have the following relation called Fannes' inequality
\be
|H(\rho)-H(\sigma)| \leq \frac{1}{e} + (\log d) ||\rho-\omega ||_1 .
\ee
\item Subadditivity of von Neumann entropy is an important property given by
\be\label{A_sa}
H(A) + H(B) \geq H(AB).
\ee
\item Finally, it follows from the subadditivity that conditioning does not increase entropy since we have
\be\label{A_cond}
H(B) \geq H(AB)-H(A) = H(B|A) .
\ee
\end{enumerate}

\end{document}